\documentclass{ws-ijmpd}

\begin{document}

%

\def\nocropmarks{\vskip5pt\phantom{cropmarks}}


%


%
\catchline{}{}{}
%

\title{A theoretical calculation of microlensing signatures caused by free-floating planets towards the
Galactic bulge}

\author{L. HAMOLLI, M. HAFIZI}

\address{Department of Physics, University of Tirana, Albania}

\author{ A.A. NUCITA}

\address{Department of Mathematics and  Physics Ennio De Giorgi and INFN, 
University of Salento, CP 193,
I-73100 Lecce, Italy}

\maketitle

\pub{Received (received date)}{Revised (revised date)}

\begin{abstract}
Free-floating planets are recently drawing a special interest of the scientific community. Gravitational microlensing 
is up to now the exclusive method for the investigation of free-floating planets, including their spatial distribution function and 
mass function. In this work, we examine the possibility that the future Euclid space-based observatory may allow to discover a substantial number of
microlensing events caused by free-floating planets. Based on latest results about the free-floating planet mass function
in the mass range $[10^{-5}, 10^{-2}]M_{\odot}$, we calculate the optical depth towards the Galactic bulge as well as the expected microlensing rate and find 
that Euclid may be able to detect hundreds to thousands of these events per month. Making use of a synthetic population, we also investigate 
the possibility of detecting parallax effect in simulated microlensing events due to free-floating planets and find a significant efficiency for 
the parallax detection that turns out to be around $30\%$.
\end{abstract}

\section{Introduction}
Gravitational microlensing is at present the only observational technique that allows the detection of extremely faint or even completely
dark objects, when their gravitational field acts as a lens to magnify background 
source stars \cite{Pac1986}. A gravitational lens is characterized by its Einstein ring radius,
\begin{equation}
R_E(M,x)=\sqrt{\frac{4GMD_s}{c^2}x(1-x)}~,\label{RE}
\end{equation}
which is the radius of the ring image formed when the observer, the lens and the source are perfectly aligned. Here $M$ is the mass of the lens and  
$x=D_l/D_s$ is the normalized lens distance, whereas $D_s$ and  $D_l$ are the source-observer and lens-observer distance, respectively.

In a microlensing event, the image separation is too small to be resolved and the observable 
feature is the variation in time of the light magnification, as due to the lens-source relative motion. 
The key parameter of the microlensing light curve is the Einstein radius crossing time given by
\begin{equation}
T_E=\frac{R_E}{v_T}~,\label{TE}
\end{equation}
where $v_T$ is the relative transverse velocity between the lens and the source.

 In the simplest case (named standard case),  when both the lens and the source can be considered as point-like objects and, additionally,
their relative motion with resepect to the observer is assumed to be linear,  the amplification of the source star follows the Paczy\'nski profile  \cite{Pac1986}
 \begin{equation}
A_s=\frac{u^2(t)+2}{u(t)\sqrt{u^2(t)+4}},
\label{As} 
\end{equation}
where
\begin{equation}
u(t)=\sqrt{u_0^2+\left(\frac{t-t_0}{t_E}\right)^2}, 
\label{ut}
\end{equation}
is the  separation between the lens  and the line of sight in units of $R_E$ and $u_0$ is the minimum separation (impact parameter) obtained at the 
moment of the peak magnification  $t_0$. The light curve obtained by equation (\ref{As}) is, evidently, symmetric around $t_0$.

 When the projected source encounters the Einstein ring of the lens, i.e. when the projected separation is  $u=1$, the source amplification takes a specific value 
defined as the threshold amplification $A_{th}=1.34$. 
For space-based telescopes, due to the absence of seeing effects, the amplification threshold may be much smaller than 1.34, with a corresponding much 
larger value for the impact parameter. For $A_{th}=1.001$, as expected  for the Euclid telescope, the maximum value of  $u$ in equation (\ref{As}) 
turns out to be $u_{max}=6.54$.

A standard microlensing light curve is described by three parameters, $t_0$, $t_E$ and $u_0$, but
only one of them, the Einstein radius crossing time, $t_E$, contains information about the lens.
As can be seen from equations (\ref{RE}) and  (\ref{TE}), the event duration is determined by three unknown parameters of the
lens: its mass $M$, the  transverse velocity $v_T$ and the distance $D_l$.
 There are several methods that have been proposed for breaking the microlensing parameter  degeneracy. Among 
them there is the parallax effects, occurring due to the motion of the Earth around the Sun and the effect of the 
relative accelerations among the observer, the lens and the source \cite{SmithMaoPacz}. These second order effects  induce small deviations in 
the light curve (with respect to the Paczy\'nski profile), which may be extremely useful to break, at least partially, the parameter degeneracy problem in microlensing observations. 

Recently, it has been reported the observation  by MOA-II of a galactic population of free objects with planetary masses, named free-floating planets 
(FFPs) \cite{Sumi2011}. Due to the intrinsic faintness of the FFPs, it is very hard to observe them directly. Hence, the gravitational microlensing method may be a suitable 
technique to detect such objects when they act as lenses for the background stars. Of course, as it will be clarified in the following, the small planetary masses imply microlensing events with 
very short time-scales.

The purpose of this paper is the investigation of the traces of  FFPs,  among other galactic lens populations, in microlensing events that might be observed serendipitously  during the planned observations of the Euclid satellite 
towards the Galactic bulge. In Section 2, we discuss the FFP mass function, their spatial distribution  and the adopted velocity
distribution. In Section 3, we review the microlensing method and its parameters in the case of Euclid observations. In Section 4, we show how  the parallax effect may help in solving, at least partially, 
the  degeneracy of parameters in the microlensing curves. 
Our main results are presented and discussed in Section 5, while in Section 6 we draw the main conclusions of this work.

\section{Planetary population}

In a recent survey of the Galactic bulge, the MOA-II collaboration \cite{Sumi2011} reported
the discovery of planetary-mass objects either very distant from their host stars (more than 100 AU away)
or entirely unbound. By analyzing the timescale distribution of all the observed microlensing events, they found a statistically significant excess of events with timescale $t<2$ days as compared
to the number of expected events from the standard Galactic model. The stellar mass-function for the standard Galactic model is generally expressed, for stars with mass $\leq 1 M_{\odot}$ and brown dwarfs (BDs), with three power laws of the form
\begin{equation}
\frac{dN}{dM} \sim M^{-\alpha_i}~~~~~\left\{\begin{array}{ll}
 {\alpha_1}=2  &  0.7M_{\odot}<M<M_{\odot} \\
 { \alpha_2}=1.3  &  0.08M_{\odot}<M<0.7M_{\odot} \\
 {\alpha_{BD}}=0.49^{+0.24}_{-0.27}  &  0.01M_{\odot}<M<0.08M_{\odot},\\
\end{array}\right.~\label{Sumi}
\end{equation}

A best-fit  procedure to the observed microlensing events due to FFPs has also allowed Sumi et al. \cite{Sumi2011}  to extend and constrain 
the power-law mass function at the low-mass regime of the FFPs 
\begin{equation}
\frac{dN}{dM}=k_{PL}M^{-\alpha_{PL}}, \hspace{0.5cm}\alpha_{PL}=1.3^{+0.3}_{-0.4}, \hspace{0.5 cm}  10^{-5}M_{\odot}<M<10^{-2}M_{\odot}.
\label{mfffp}
\end{equation}
The derived number of planetary mass objects per star turns out to be very large, 
although rather  poorly constrained: $N_{PL}=5.5^{+18.1}_{-4.3}$, mainly due to the poor precision of the lens mass estimate  
below $10^{-4}M_{\odot}$.   

The abrupt change from $\alpha_{BD}=0.49$ to $\alpha_{PL}=1.3$ favors the idea of a separate population, whose
formation process is different from that of stars and BDs. These objects may have formed in proto-planetary 
disks and subsequently scattered into unbound or very distant orbits, becoming  FFPs.

From the likelihood contours of the power-law indices found in the brown dwarf and  planetary-mass
regime \cite{Sumi2011}, we get a useful correlation between $\alpha_{BD}$  and $\alpha_{PL}$
\begin{equation}
\alpha_{BD} =1.7-\alpha_{PL}~ \label{alfa3}.
\end{equation}

Regarding the spatial distribution of the FFPs, we assume that they are distributed as the stars in the 
Milky way \cite {Gilmore1989,Depaolis2001,Mimoza}. Hence, we considered the following density distributions:    \\    
1. exponential thin disk, 
\begin{equation}
\rho(R,z) =\rho_0^{\rm D_{thin}}(M) ~e^{-|z|/H}~e^{-(R-R_0)/h},\label{diskthin}
\end{equation}
in cylindrical coordinates $R$  (the galactocentric distance in the galactic plane) and $z$ (the distance from the galactic plane). The scale parameters are $H\sim 0.30$ kpc, $h\sim 3.5$ kpc; $R_0=8.5$ kpc is the local galactocentric distance.\\
2. exponential thick disk,
\begin{equation}
\rho(R,z) =\rho_0^{\rm D_{thick}}(M) ~e^{-|z|/H}~e^{-(R-R_0)/h},\label{diskthick}
\end{equation}
with $H\sim 1$ kpc, $h\sim 3.5$ kpc and $R_0=8.5$ kpc.\\ 
3. triaxial bulge \cite {Depaolis2001,Mimoza,Dwek1995} 
\begin{equation}
\rho(x,y,z) =\rho_0^{\rm Bulge}(M)e^{-s^2/2}~,~~~{\rm
with}~~~ s^4=(x^2/a^2+y^2/b^2)^2+z^4/c^4~, \label{bulge}
\end{equation}
where $ a=1.49$ kpc, $b=0.58$ kpc, $c=0.40kp$c.

For the  FFP velocity distribution, we assume for each coordinate the Maxwellian distribution \cite {HanGould1995,HanGould1996}   
\begin{equation}
f(v_i) \propto \exp^{-\frac{(v_i-\overline{v}_i)^2}{2\sigma^2_i}},~~~~i\in \{x, y, z \},  
\end{equation}
where the coordinates  $ (x, y, z)$ have their origin at the galactic center and the x and z-axes point to the Sun and the north Galactic pole, respectively. 
We are interested only to the  perpendicular velocity with respect to the line of sight, namely to y and z components.
For lenses in the Galactic  bulge we use the mean velocity components  $\overline{v}_y=\overline{v}_z=0$, with dispersion  $\sigma_y=\sigma_z=100$ km/s;
for lenses in the Galactic disk we use  the mean velocity components $\overline{v}_y=220$ km/s, $\overline{v}_z=0$,  with dispersion velocity
$\sigma_y=\sigma_z=30$ km/s for the thin disk and  $\sigma_y=\sigma_z=50$ km/s for the  thick disk.

\section{Microlensing events towards the Galactic bulge}

Several  microlensing surveys with relatively high image sampling  have been undertaken until now towards the Galactic bulge by the MOA (Microlensing
Observations in Astrophysics) Collaboration \cite{moa} and the OGLE (Optical Gravitational Lensing Experiment) Collaboration \cite{ogle2006}  (to cite only some of them), with the aim 
of searching for MACHOs (Massive Astrophysical Compact Halo Objects) and exoplanets. These surveys (undertaken since about two decades) have allowed the 
detection of several thousands of microlensing events, most of which are due to self-lensing (stars either in the Galactic disk and bulge). Ground-based 
observations may detect low-mass lenses (short time duration events) only with great difficulties, so to search for lens masses below   
$ 0.01 M_{\odot}$, as for FFPs,
space-based observations are needed. At present, there are two space-based 
missions which are planned for detecting microlensing events towards the Galactic bulge: the
Wide-Field Infrared Survey Telescope (WFIRST) and Euclid.

Euclid is a Medium Class mission of the ESA (European Space Agency), which is scheduled to be launched in 2017. For ten months, not necessarily 
consecutive \cite{Euclid}, it will perform microlensing observations towards the  Galactic bulge.
The galactic coordinates of the Euclid line of sight are $b=-1.7 ^{\circ}$,  $l=1.1^{\circ}$, the distance of observation can be considered  $D_s=(7-10)$ kpc, with mean value at $D_s=8.5$ kpc and the observing image rate (cadence) is expected to be about $20$ min.

In order to study the expected  microlensing events that may be detected by the Euclid observatory we start by evaluating the microlensing optical depth and the event rate. 
The microlensing  optical  depth  is  defined  as  the  probability  that  at  any  time  a  random  star  is 
magnified more  than the  threshold amplification  $A_{th}=1.34$  by a lens belonging to a given  population of lenses. It is given by  (see e.g. \cite {Griest,Jetzer2002}) 
\begin{equation}
\tau=\int_0^{D_s} n(D_{l})\pi R_E^2d D_{l} = \frac{4\pi
GD_s^2}{c^2}\int_0^1 \rho(M,x) x (1-x)dx~,\label{taoe}
\end{equation}
where $\rho(M,x)$  is the mass density of the lens population; $x=D_l/D_s$.

The microlensing rate is the number of events per unit time and per monitored star due to the lens population. It is given by (see e.g. \cite{Griest,Jetzer2002}) 
\begin{equation}
\Gamma=\int\frac{ n(x) f({\bf v}_l - {\bf v}_t) f({\bf
v}_s) dx d{\bf v}_l d{\bf v}_s}{dt}, \label{gammae}
\end{equation}
where ${\bf v}_l$, ${\bf v}_s$ and ${\bf v}_t$ are the lens, the source and the microlensing tube two-velocities in the plane
transverse to the line of sight. 
The velocity distribution functions $ f({\bf v}_l)$  and  $ f({\bf v}_s)$  are assumed to have Maxwellian forms \cite {HanGould1995,HanGould1996}, with one-dimensional dispersion velocities 
different for each lens and source population. The tube velocity  is  given by 
\begin{equation}
v_t^2 (x) = (1-x)^2 v_{\odot}^2 + x^2 v_s^2 + 2x(1-x)v_{\odot}v_s
\cos\theta~,
\end{equation}
where ${\bf v}_{\odot}$ is the local velocity transverse to the
line of sight and $\theta$ is the angle between ${\bf v}_{\odot}$
and ${\bf v}_s$.
                  
In the case of observations towards the Galactic bulge, the source stars are mostly bulge stars which are distributed following, as usual, the same  triaxial mass density model as given 
in eq.  (\ref{bulge}) with $\rho_0^{\rm Bulge}=M_b/(8\pi abc)$, 
where $M_b \simeq 2 \times 10^{10}~M_{\odot}$, $ a=1.49$ kpc, $b=0.58$ kpc and $c=0.40$
kpc.

The limiting line flux of Euclid Telescope is estimated to be  $ F_ l=3\times10^{-19} J s^{-1}m^{-2}$, \cite{Euclid} whereas the flux of a Sun-like star situated at the Galactic center
is $F_{\odot}=4.44\times10^{-16} J s^{-1}m^{-2}$. Based on the mass-luminosity relation $ \frac{L}{L_{\odot}}=(\frac{M}{M_{\odot}})^{2.4}$ for low-mass stars
($M<0.8M_{\odot}$), it can be directly 
shown that the telescope can observe all bulge stars. 

The mean mass for bulge stars is $<M>=0.27M_{\odot}$, found by using the Salpeter mass 
function\cite{salpeter} $\frac{dN}{dM}\sim M^{-2.4}$; the Euclid's field of view is $0.54$ square degree, hence the number of source stars in Euclid microlensing observations will be  $N_{ED}=2.3\times 10^8 $. 
This number has to be multiplied by the microlensing rate (\ref{gammae}) and the time of observation (in the following we take $t_{\rm obs}=1$ month) to get an estimate of the number of microlensing event  that we expect to be detectable by the Euclid telescope. \\

\section{Parallax effect}

The parallax effect due to the motion of the Earth around the Sun may leave in microlensing events some  observable feature which can be used to break the 
degeneracy of microlensing parameters, or at least to constrain the microlensing parameter space. Here, we are focusing on the investigation of the parallax traces left on 
microlensing events that will possibly detected by the Euclid telescope.
In order to estimate the parallax effects on the microlensing
light curves  we make use of the following useful geometrical relations \cite{Dom1998}
\begin{equation}
\begin{array}{l}
A_p=\frac{u^2(t)+2}{u(t)\sqrt{u^2(t)+4}} \\\\
 u^2(t)=p^2(t)+d^2(t)~\\\\
p(t)=p_0(t) +\cos{\psi}[x_1(t)-x_1(t_0)]+\sin{\psi}[x_2(t)-x_2(t_0)]\\\\
d(t)=d_0 -\sin{\psi}[x_1(t)-x_1(t_0)]+\cos{\psi}[x_2(t)-x_2(t_0)]\\\\
x_1(t)=\rho[-\sin{\chi}\cos{\phi}(\cos{\xi(t)}-\epsilon)-\sin{\chi}\sin{\phi}\sqrt{1-\epsilon^2}\sin{\xi(t)}]\\\\  
x_2(t)=\rho[-\sin{\phi}(\cos{\xi(t)}-\epsilon)+\cos{\phi}\sqrt{1-\epsilon^2}\sin{\xi(t)}]\\\\ 
\rho=\frac{a_{\oplus}(1-x)}{R_E}~~~~~~p_0(t)=\frac{(t-t_0)}{T_E}~~~~~~d_0=u_0  \\\\
\end{array}~\label {Ap},
\end{equation}
where $\xi(t)$ is implicitly given by
\begin{equation}
t=\sqrt{\frac{a_{\oplus}^3}{GM_{\odot}}(\xi-\epsilon\sin{\xi})}.                                                
\end{equation}
Here,  $a_{\oplus}$ is the semi-major axis of the Earth orbit around the Sun,  $\epsilon=0.0167$ is the Earth orbit eccentricity and 
$\rho$ is the length of the semi-major axis projected onto the lens plane
measured in Einstein radii.  The position of the source stars is  characterized by the parameters $\phi,\chi$ and $\psi$ in the relations (\ref{Ap}) which  give, respectively, the longitude  measured in the ecliptic
 plane from perihelion towards the Earth motion, the latitude  measured from the ecliptic plane towards the northern point of the ecliptic and the rotation angle 
 in the lens plane which describes the relative orientation of velocity $v_T $   to the sun-earth system. 
 We find that the deviations on the microlensing light curve due to the parallax effect depend substantially on the Earth 
position in its orbit  at the time of the maximum amplification and get the largest value when $\xi_0=165^{\circ}$ that happens in June.
 
In the following calculations, we have assumed that the Euclid satellite is in the best position in its orbit in order to maximize the parallax effect on the microlensing event light curves. 
Using  the usual transformation relations  between coordinate systems, we find the following values for the Euclid's line of sight towards the Galactic bulge: $\phi\simeq 167.8^{\circ}$             
 and $\chi\simeq-5.4^{\circ}$. 

To the aim of estimating the number of events for which the parallax feature affects the microlensing light curves as observed by Euclid, 
we make use of Monte Carlo numerical simulations by generating microlensing events towards the field of the sky planned to be observed by the Euclid telescope. 
Here, we briefly describe the adopted strategy. In particular, we draw\\
a)  lens distances $D_l$, based on the above written disk or bulge spatial distributions. We always consider the source as being located in the Galactic bulge, so we fixed $D_s=8.5$ kpc for all events;\\
b)  the relative transverse velocity from the velocity distribution;\\ 
c)  the impact parameter randomly distributed on a uniform interval  [0,6.54]. As already anticipated, the Euclid amplification threshold  is planned to be $A_{th}=1.001$; \\ 
d)  the lens mass that follows the mass function distribution in eq. (\ref{mfffp}).\\ 
In each case we choose the same position of the Earth, $\xi_0=165^{\circ}$, at the time $t_0$  of the light curve  peak amplification.

\begin{figure}[htbp]
\vspace{10cm} \includegraphics{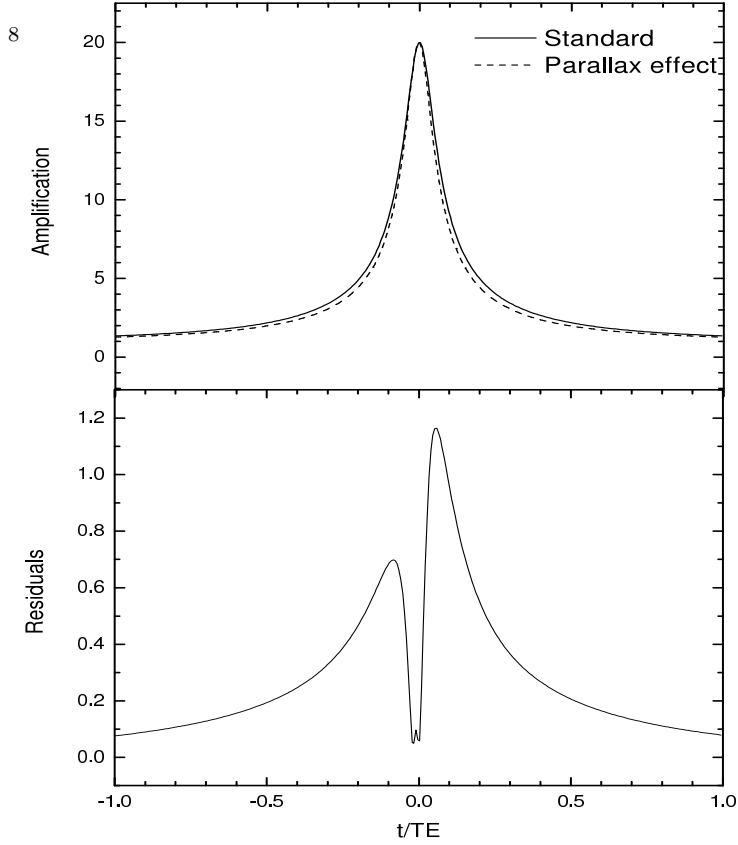} \caption{Upper panel: the standard Paczy\'nski light curve (continuous line) and the parallax curve (dashed line)
for a planet with mass $10^{-3}M_{\odot}$ at distance  $D_l=4.5$ kpc from Earth are shown.  The residual curve between the two curves is also shown (bottom panel).} \label{22222}
\end{figure}

The parallax effect is estimated by calculating the residuals between the light curve $A_p(t)$ containing the parallax effect from eqs. (\ref{Ap}) and the corresponding 
standard curve $A_s(t)$ (\ref{As}), i.e. $Res=|A_s(t)-A_p(t)|$. As an example, in Fig.\ref{22222} we show the standard curve, the
parallax curve and residuals in the case of a free-floating planet with mass $10^{-3}M_{\odot}$  at distance $D_l=4.5$ kpc from Earth. As one can see, in this case the residuals are up to $\simeq 12\%$.

\section{Results}

We have calculated the microlensing optical depth from eq. (\ref{taoe}) and the microlensing rate from eq.  (\ref{gammae}) for all the lens populations towards the Galactic bulge: FFPs, BDs and 
stars distributed in the thin disk, thick disk and the Galactic  bulge. 
For their spatial distribution we use eqs. (\ref{diskthin}), (\ref{diskthick}),  (\ref{bulge}), coupled with the mass functions (\ref{Sumi}) and  (\ref{mfffp}). For stars we assume
the Salpeter mass function $\frac{dN}{dM}\sim M^{-2.4}$, while the relation
 between $\alpha_{BD}$  and $\alpha_{PL}$ is defined by equation (\ref{alfa3}).

Brown dwarfs are faint objects distinguished only in Sun surroundings.  Recently, the ratio $R$ (in the Solar Neighbourhood) between the number of stars with mass in
the range $[0.08, 1]M_{\odot}$ and BDs in the range $[0.03,0.08]M_{\odot}$, has been estimated by \cite{Parravano}. From Table 1 in  \cite{Parravano}  we adopt a mean value of $R\simeq 5.1$  and assume that
it applies to the whole Galaxy.

In Table \ref{tao}, we show the results of our calculations of the optical depth for FFPs and BDs considered as lenses. We perform separate calculations for each structure of the 
galaxy  (bulge, thin disk and thick disk) and different values of $\alpha_{PL}$. The number of FFPs per star is chosen following Sumi et al. \cite{Sumi2011}: the 
lowest value $N_{PL}=1.2$, mid value $N_{PL}=5.5$ and highest value 
$N_{PL}=23.6$. The optical depth yielded by stars is not dependent on $\alpha_{PL}$: our calculations give $2.59\times 10^{-6}$ for bulge stars, 
$4.26\times 10^{-7}$ for thin disk stars and $2.42\times 10^{-7}$ for thick disk stars, respectively.\\

\begin{table}[htbp]
\ttbl{30pc}{Optical depth for FFPs and BDs, distributed in bulge, thin and thick disk, for different values of $\alpha_{PL}$ and for fixed values $23.6$, $5.5$ and $1.2$ of the number of FFPs per star.}
\centering
\scriptsize
\begin{tabular}{|c|c|c|c|c|c|c|c|c|c|c|c|c|}
 \hline
 &\multicolumn{3}{|c|}{$N_{PL}=1.2$}&\multicolumn{3}{|c|}{$N_{PL}=5.5$}&\multicolumn{3}{|c|}{$N_{PL}=23.6$}&\multicolumn{3}{|c|}{$BD$} \\
 &\multicolumn{3}{|c|}{$(\times 10^{-9})$}&\multicolumn{3}{|c|}{$(\times 10^{-8})$}&\multicolumn{3}{|c|}{$(\times 10^{-8})$}&\multicolumn{3}{|c|}{$(\times 10^{-8})$} \\
\hline$\alpha_{PL}$&$Blg$&$D_{thn}$&$D_{thc}$&$Blg$&$D_{thn}$&$D_{thc}$&$Blg$&$D_{thn}$&$D_{thc}$&$Blg$&$D_{thn}$&$D_{thc}$ \\
\hline$0.9$&$22.1$&$3.52$&$2.00$&$10.1$&$1.61$&$0.92$&$43.4$&$6.92$&$3.93$&$9.95$&$1.59$&$0.90$\\
\hline$1.0$&$17.6$&$2.80$&$1.59$&$8.04$&$1.28$&$0.73$&$34.5$&$5.51$&$3.12$&$10.0$&$1.60$&$0.91$\\
\hline$1.1$&$13.5$&$2.16$&$1.22$&$6.19$&$0.99$&$0.56$&$26.5$&$4.24$&$2.40$&$10.1$&$1.61$&$0.92$\\
\hline$1.2$&$10.1$&$1.62$&$0.92$&$4.64$&$0.74$&$0.42$&$19.3$&$3.19$&$1.81$&$10.1$&$1.63$&$0.92$\\
\hline$1.3$&$7.43$&$1.18$&$0.67$&$3.40$&$0.54$&$0.31$&$14.6$&$2.32$&$1.32$&$10.3$&$1.64$&$0.93$\\
\hline$1.4$&$5.36$&$0.86$&$0.49$&$2.46$&$0.39$&$0.22$&$10.5$&$1.68$&$0.95$&$10.3$&$1.65$&$0.93$\\
\hline$1.5$&$3.84$&$0.61$&$0.35$&$1.76$&$0.28$&$0.16$&$7.54$&$1.21$&$0.68$&$10.4$&$1.67$&$0.94$\\
\hline$1.6$&$2.74$&$0.48$&$0.25$&$1.26$&$0.20$&$0.12$&$5.40$&$0.86$&$0.49$&$10.4$&$1.68$&$0.95$\\
 \hline
\end{tabular}
\normalsize
\label{tao}
\end{table}
We remark that the contribution of the bulge populations in the optical depth is the most important one. 

In Table \ref{gama}, we show the results of our calculations for the microlensing rate (the probability to have a microlensing event per star during the observation time of one month),
 in the same conditions as above. The microlensing rate by stars is of course independent on $\alpha_{PL}$, its value being  
$2.41\times 10^{-6}$ for bulge stars, 
$1.27\times 10^{-7}$ for thin disk stars and $1.69\times 10^{-7}$ for thick disk stars, respectively.

\begin{table}[htbp]
\ttbl{30pc}{Microlensing rate (the probability to have a microlensing event per star during the observation time of one month)  for FFPs and BDs, distributed in bulge, thin and thick disk, for different values of $\alpha_{PL}$ and for fixed values $N_{PL}=1.2, 5.5$ and $23.6$ of the number of FFPs per star.}
\centering
\scriptsize
\begin{tabular}{|c|c|c|c|c|c|c|c|c|c|c|c|c|}
 \hline
 &\multicolumn{3}{|c|}{$N_{PL}=1.2$}&\multicolumn{3}{|c|}{$N_{PL}=5.5$}&\multicolumn{3}{|c|}{$N_{PL}=23.6$}&\multicolumn{3}{|c|}{$BD$} \\
 &\multicolumn{3}{|c|}{$(\times 10^{-8})$}&\multicolumn{3}{|c|}{$(\times 10^{-8})$}&\multicolumn{3}{|c|}{$(\times 10^{-7})$}&\multicolumn{3}{|c|}{$(\times 10^{-8})$} \\
\hline$\alpha_{PL}$&$Blg$&$D_{thn}$&$D_{thc}$&$Blg$&$D_{thn}$&$D_{thc}$&$Blg$&$D_{thn}$&$D_{thc}$&$Blg$&$D_{thn}$&$D_{thc}$ \\
\hline$0.9$&$21.7$&$1.14$&$0.91$&$93.3$&$5.23$&$4.18$&$42.6$&$2.25$&$1.79$&$23.3$&$1.24$&$1.66$\\
\hline$1.0$&$18.5$&$0.98$&$0.78$&$84.7$&$4.47$&$3.56$&$36.3$&$1.92$&$1.53$&$23.6$&$1.25$&$1.66$\\
\hline$1.1$&$15.5$&$0.82$&$0.65$&$71.0$&$3.75$&$2.99$&$30.5$&$1.61$&$1.28$&$23.7$&$1.25$&$1.67$\\
\hline$1.2$&$12.9$&$0.68$&$0.54$&$59.0$&$3.12$&$2.49$&$25.3$&$1.34$&$1.07$&$23.7$&$1.26$&$1.67$\\
\hline$1.3$&$10.7$&$0.56$&$0.45$&$48.8$&$2.59$&$2.05$&$20.9$&$1.11$&$0.88$&$23.9$&$1.26$&$1.68$\\
\hline$1.4$&$8.95$&$0.47$&$0.37$&$40.6$&$2.14$&$1.71$&$17.4$&$0.92$&$0.73$&$24.0$&$1.27$&$1.69$\\
\hline$1.5$&$7.43$&$0.39$&$0.31$&$34.1$&$1.80$&$1.43$&$14.6$&$0.78$&$0.62$&$24.1$&$1.27$&$1.69$\\
\hline$1.6$&$6.33$&$0.34$&$0.27$&$29.0$&$1.54$&$1.22$&$12.5$&$0.66$&$0.53$&$24.2$&$1.28$&$1.70$\\
 \hline
\end{tabular}
\label{gama}
\end{table}
The dominant contribution in the microlensing rate is again that of the bulge lens populations.
 We then estimate the number of microlensing events expected to be detectable by the Euclid telescope (taking  $A_{th}=1.001$ and therefore $u_{max}=6.54$) multiplying  the microlensing event rate by the number of source stars in the Euclid field of view $N_{ED}=2.3\times 10^8 $ and by the observation time duration.

In Table \ref{nevents},  we show the results of our calculations for the estimated number of microlensing events per month, where FFPs and BDs are considered as lenses. We perform separate calculations 
for each structure of the 
galaxy: bulge, thin disk and thick disk. The number of microlensing events per month produced by stars is not dependent on $\alpha_{PL}$, its value is $3657$ for bulge stars, 
$193$ for thin disk stars and $265$ for thick disk stars.\\

\begin{table}[htbp]
\ttbl{30pc}{Number of microlensing events detectable by the Euclid telescope in one month of observation towards the Galactic bulge  for the different lens  populations  and for different values of $\alpha_{PL}$.}
\centering
\scriptsize
\begin{tabular}{|c|c|c|c|c|c|c|c|c|c|c|c|c|}
 \hline
 &\multicolumn{3}{|c|}{$N_{PL}=1.2$}&
\multicolumn{3}{|c|}{$N_{PL}=5.5$}&
\multicolumn{3}{|c|}{$N_{PL}=23.6$}&
\multicolumn{3}{|c|}{$BD$} \\
\hline$\alpha_{PL}$&$Blg$&$D_{thn}$&$D_{thc}$&$Blg$&$D_{thn}$&$D_{thc}$&$Blg$&$D_{thn}$&$D_{thc}$&$Blg$&$D_{thn}$&$D_{thc}$ \\
\hline$0.9$&$328$&$17$&$14$&$1505$&$79$&$63$&$6457$&$340$&$272$&$354$&$19$&$25$\\
\hline$1.0$&$280$&$15$&$12$&$1284$&$68$&$54$&$5507$&$291$&$232$&$358$&$19$&$25$\\
\hline$1.1$&$235$&$12$&$10$&$1077$&$57$&$45$&$4603$&$244$&$195$&$359$&$19$&$25$\\
\hline$1.2$&$195$&$10$&$8$&$894$&$47$&$38$&$3838$&$203$&$162$&$359$&$19$&$25$\\
\hline$1.3$&$161$&$9$&$7$&$740$&$39$&$31$&$3175$&$168$&$134$&$362$&$19$&$26$\\
\hline$1.4$&$134$&$7$&$6$&$615$&$33$&$26$&$2638$&$139$&$111$&$363$&$19$&$26$\\
\hline$1.5$&$113$&$6$&$5$&$516$&$27$&$22$&$2215$&$117$&$93$&$365$&$19$&$26$\\
\hline$1.6$&$96$&$5$&$4$&$440$&$23$&$19$&$1888$&$100$&$80$&$366$&$19$&$26$\\
 \hline
\end{tabular}
\normalsize
\label{nevents}
\end{table}
The bulge population contribution is the most important one also in this case. 

\begin{figure}[htbp]
\vspace{10cm} \includegraphics{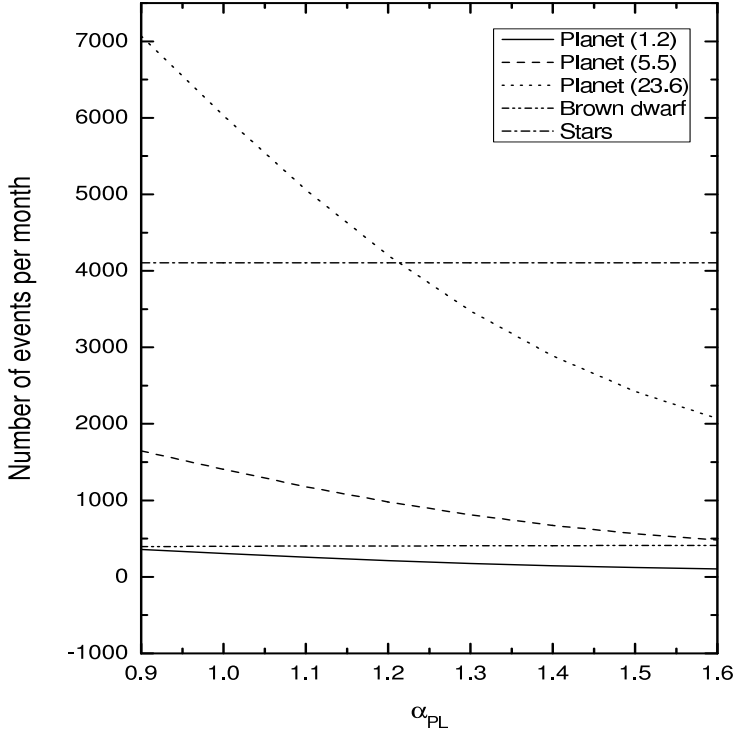} \caption{The total microlensing  event number due to BDs (dot-dot-dashed line), stars  (dot-dashed line) and FFPs  expected to be detectable in one month of observation towards the Galactic bulge by the Euclid telescope as a 
function of $\alpha_{PL}$. The three curves for the FFP contribution are drawn assuming $N_{PL}=5.5$ (dashed line), $N_{PL}=23.6$ (dotted line),
$N_{PL}=1.2$ (continuous line).} \label{numri}
\end{figure}

In Fig.\ref{numri} we present our estimations for the total number of microlensing events due to BDs and FFPs per month, for different values of $\alpha_{PL}$.

By numerical simulations we produce a large number of microlensing events caused by the population of the FFPs.
We assume that a microlensing event can be detected if in its light curve there are at least 8 points in which the amplification is bigger than the threshold amplification $A_{th}=1.001$. The photometric error in this case is 0.1\%  
\begin{equation}
A_{th}F-F=F( A_{th}-1)=\Delta{F} {\Rightarrow} ( A_{th}-1)=\frac{\Delta{F}}{F}=1.001-1=0.001.                         
\end{equation}
In the case of Euclid telescope, the expected curve will contain points determined every 20 minutes, that means that any detectable 
event has to have a duration larger than 2.67 hours.\\
For estimating the parallax effect on the observed light curves, we consider only those containing at least 8 points with $Res>0.001$ inside Einstein ring
\begin{equation}
|A_s(t)F-A_p(t)F|>\Delta{F}{\Rightarrow} |A_s(t)-A_p(t)|>\frac{\Delta{F}}{F}{\Rightarrow}Res>0.001.                          
\end{equation}
We retain all synthetic events with residuals fulfilling the above-mentioned condition. Therefore, the efficiency for parallax effect detection is given by the ratio between the number of these events and the total number of  detectable events. 

\begin{figure}[htbp]
\vspace{10cm} \includegraphics{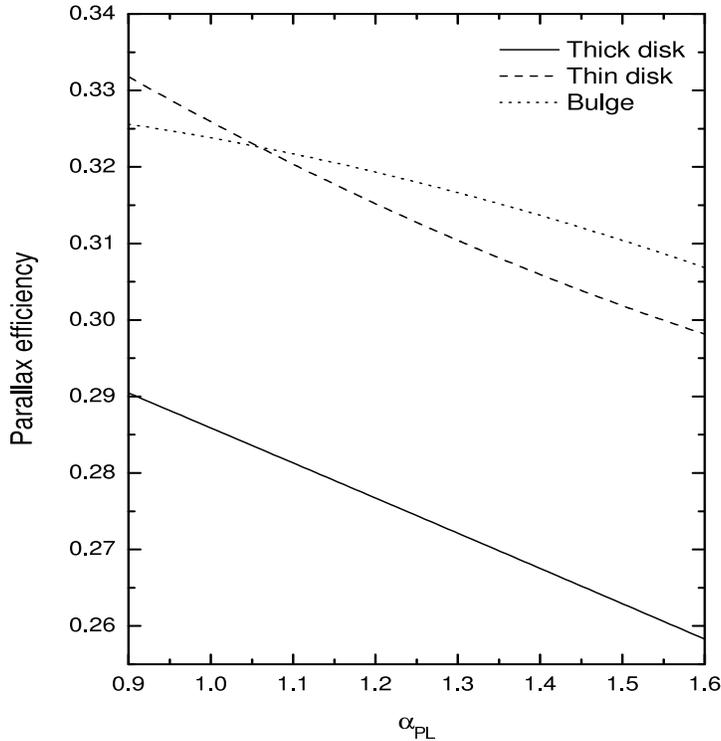} \caption{Parallax efficiency caused by free-floating planets as a function of $\alpha_{PL}$  for the three different distributions of FFPs:
 bulge FFPs (dotted line), thin disk FFPs (dashed line) and thick disk FFPs (continuous line).} \label{eficienca1}
\end{figure}

In Fig.\ref{eficienca1} we show  our results for the parallax efficiency in microlensing events caused by FFPs and expected to be detectable by Euclid are shown. 
We consider three separate distributions, bulge, thin and thick disk lenses and show the parallax efficiency with respect to the value of $\alpha_{PL}$.\\

 For example, as one can see, in the  case of $\alpha_{PL}=1.3$  the parallax efficiency caused by bulge FFPs is $ 32\%$, by thin disk FFPs is $31\%$ and by thick disk FFPs is $27\%$. 
So, approximately  in $30\%$ of the detectable events, the parallax effect due to the Earth motion may be detectable by the Euclid telescope, allowing to partially resolve the parameter degeneracy problem  in this kind of observations  and constrain  the distance to the FFPs. This should allow to estimate not only the number of FFPs throughout the Milky Way, but also their spatial distribution.

\section{Conclusions}
In this paper we investigate the possible observation of free-floating planets (in addition to normal stars and brown dwarfs) towards the Galactic bulge by the future Euclid space-based observatory, via 
detection of microlensing light curves. These events, either considered statistically or individually, are an important 
base of knowledge to better characterize the galactic populations of objects in addition to  normal stars.

For the calculation of the optical depth and the microlensing rate for brown
dwarfs we assume that  they are distributed like stars, spatially and in the velocity space.  
This should also be true for free-floating planets, based on the idea that these objects are most likely formed in proto-planetary
disks and subsequently scattered into unbound or very distant orbits. The number of these objects per star is poorly 
constrained, as is also the slope of their mass distribution.

We find that the optical depth and the microlensing rate for FFPs towards the Galactic bulge are much smaller 
than for stars, but slightly higher than for brown dwarfs. The highest contribution  for the three object populations always comes from  bulge objects.
The theoretical optical depth and microlensing rate depend on the  power law index of the FFP mass function, hence the corresponding 
observed values can be considered as sources of information for the still largely unknown mass function of brown dwarfs and FFPs. 

By theoretical calculations we predict that a considerably large number of microlensing events produced by free-floating 
planets towards the  Galactic bulge are potentially observable by the Euclid satellite. 
We also take into account the deviations in the microlensing light curves due to FFPs induced by the Earth parallax effect. We find that these  deviations depend substantially on the
Earth position in its orbit around the Sun at the time of the event maximum amplification and get the largest value in June.
By numerical simulations we also find that the efficiency (that is the ratio between the number of events due FFPs that fulfill equation (18) with respect to the total number of detectable events) of detecting the 
Earth parallax effect in the light curves due to FFPs is potentially interesting since the parallax effect turns out to be detectable in about  1/3 of all observable events (see Fig.\ref{eficienca1}). 
We emphasize that the observation of this effect may allow to constrain the FFP distances, which is a fundamental information necessary to investigate 
 how FFPs are distributed throughout the Milky Way. This, in turn, is an important issue in order to establish their origin.

As a final remark we caution that the short time-scale microlensing features, such as those expected due to the  Earth parallax, may be confused due to the so-called red-noise effect. 
Indeed, photometric 
observations are generally affected by the presence of the Earth atmosphere that is a source of correlated noise. A way to circumvent this problem is to use space-based telescopes, an opportunity that has clearly many advantages. 
However, the improved sensitivity of space-based observations have unveiled a new source of noise related to the intrinsic stellar variability that induces the red-noise, connected to the 
correlated time-series. This effect has been studied in connection to the transit technique when searching for exoplanets as observed by space telescopes such as CoRoT and Kepler  \cite{pont,carpano}. The detailed study of this effect in connection to the microlensing lighcurves, in particular in connection to the searches for free-floating planets,  is left to a following work.\\

We would like to thank  the  colleagues who have discussed the subject of this paper with us and particularly Francesco De Paolis for guidance and useful comments.

\end{document}